\documentclass[12pt]{article}
\usepackage[english]{babel}
\usepackage{graphicx,subfigure}
\usepackage{amssymb,amsthm,amsmath,amsfonts}
\usepackage{float}
\usepackage[colorlinks=true,linkcolor=blue,citecolor=red, urlcolor=green]{hyperref} 
\usepackage{cite}

\begin{document}

\title{A thermodynamical suspension model\\ for blood}

\author{M.~Gorgone, C.F.~Munaf\`o, A.~Palumbo, P.~Rogolino\\
\ \\
{\footnotesize Department of Mathematical and Computer Sciences,}\\
{\footnotesize Physical Sciences and Earth Sciences, University of Messina}\\
{\footnotesize Viale F. Stagno d'Alcontres 31, 98166 Messina, Italy}\\
{\footnotesize mgorgone@unime.it; carmunafo@unime.it; apalumbo@unime.it; progolino@unime.it}
}

\date{Published in \emph{Meccanica} \textbf{59}, 1671--1683 (2024).}

\maketitle

\begin{abstract}A complete thermodynamical analysis for a blood model, based on mixture theory, is performed. The model is developed considering the blood as a suspension of red blood cells (solid component) in the plasma (fluid component), and taking into account the temperature effects. Furthermore, two independent scalar internal variables are introduced accounting for {additional dissipative effects}.
Using Clausius-Duhem inequality,  the general thermodynamic restrictions and residual dissipation inequality are derived. The thermodynamic admissibility with the second law of thermodynamics
is assessed by means of the extended Coleman-Noll procedure; in one space dimension we exhibit a solution of all the thermodynamical constraints.
\end{abstract}


\section{Introduction}\label{sec1}
Roughly speaking, blood is a fluid with a complex chemical composition. It can be considered as a suspension of cells in an aqueous solution of electrolytes and 
non-electrolytes containing many enzymes and
hormones, and carryiing oxygen and carbon dioxide between lungs and tissue cells. 
Through a centrifuge process, blood can be separated into plasma and cells \cite{Fung}, the latter consisting mainly of red blood cells (RBCs),  or erythrocytes, various white blood cells and platelets. 

Although blood exhibits some features of non-Newtonian fluids, at a first approximation it may be thought of as a homogeneous Newtonian fluid flowing in large blood vessels  \cite{Donald}.
Its viscosity varies with the percentage of the total volume of blood occupied by the cells, the hematocrit, and with the temperature \cite{Cinar, KumShaMek, TrSha}.

The basic principles of fluid mechanics and heat transport can be used in
the analysis of blood flow to describe its behavior under different conditions.
Many studies are concerned with blood fluid dynamics because of its relevant applications in biomedical engineering (see, for instance, \cite{PriGur, TiwCha, TrSha}). 

The onset of many cardiovascular diseases, determining serious pathologies,
or even death, is often related to altered hemodynamic behavior in the circulatory system \cite{TsiSteMav}. For example, the aggregation of RBCs and a consequent increased viscosity alter the transport properties of blood. The deformability of RBCs is also associated to migration through the vessel wall. Other diseases can determine a reduction in the deformability of erythrocytes and an increase of hematocrit and thus viscosity.

For the therapy of cardiovascular diseases, it is often necessary to assess the
blood flow in vessels.  Many studies devoted to the nature of blood flow investigated the relationship between different parameters, such as viscosity and hematocrit.
In turn, vasodilatation (vasoconstriction, respectively)  processes involve a variation of viscosity which becomes lower (higher, respectively).

In the literature, different models of blood flow have been developed \cite{MasKimAnt, Mass2008, WuAubAntMas, MasKir}. 
In other contributions \cite{Ste, ArmPin, AnaMas}, some models suitable to describe the rheological features of blood, accounting for deformability and aggregation of red blood cells,  have been proposed.
In \cite{TsiSteMav}, a model has been derived using a generalized  bracket approach of nonequlibrium thermodynamics that takes into account reversible formation
and separation or accumulation of red blood cells involving a kinetic model. 

Experimental data suggest that changes in blood viscosity depend on the temperature, whereupon many researchers proposed  models with
a temperature dependent viscosity \cite{Cinar, KumShaMek}. Moreover,  the effects of variable viscosity,  hematocrit-dependent, on blood flow through an artery with stenosis has been investigated \cite{TrSha}.

Experimental measures show that viscosity decreases as temperature increases.   The rheological properties of blood, such as the dependence of viscosity on shear velocity and temperature, are mainly established by the action of erythrocytes to form and destroy aggregates, deform and allocate in the blood flow.
Anyway,  increasing the velocity gradient, the erythrocytes separate from each other and the viscosity decreases until it becomes almost constant.
An increase in hematocrit promotes the formation of aggregates and consequently  blood viscosity becomes higher. The influence of temperature on  aggregate generation 
is also evident.
As the temperature becomes higher,  the aggregates are destroyed by thermal motion and the viscosity decreases.

The blood mechanical properties are strongly influenced by the behavior of red blood cells due to their high volume fraction (hematocrit).

From a mathematical point of view, the flow of a RBCs-plasma suspension can be modeled within the framework of mixture theory, introduced 
in the scheme of continuum mechanics by Truesdell \cite{True}; in the last decades, this theory has been widely used in several applications involving both reacting and 
non-reacting immiscible mixtures, 
soft tissues growth, as well as blood flow \cite{WuAubMasKimAnt, MasKimAnt, KirMas}.

Several applications of mixture theory can be found in \cite{BowGar, Mul1, GurVar, Bow, LiuMul, GouRug, BotDre, FraPalRog, CimGorOliPac}, where models with different degrees of detail have been investigated.

The models considering as fundamental fields the mass densities,  temperatures and velocities of both constituents provide the most detailed description of the thermodynamics of the mixtures. These are quite important, 
for example, in  theories of plasma, where  different plasma constituents  may exhibit different temperatures on time scales whose magnitude is comparable with the transport process times. This problem has been 
first considered in \cite{BowGar} and, more recently, in \cite{GouRug}.
Furthermore, in some physical instances of complex continua modeled as fluid mixtures, internal state variables can be introduced as additional fields (see, for instance, \cite{FraPalRog}). 

As noted above, the temperature has a considerable influence on the formation of aggregates and therefore on the viscosity, thus its contribute is fundamental  in the study of blood flow.

Moving from the purely mechanical model introduced by Massoudi et al. \cite{MasKimAnt}, where the plasma is modeled as a Newtonian fluid and RBCs as an anisotropic nonlinear density--gradient fluid, whereas the effects of temperature are not taken into account, here we propose a thermomechanical model of blood considered as a suspension. In particular, we assume that plasma and RBCs may have two different temperatures, and introduce two internal variables for modeling additional dissipative effects. The thermodynamical compatibility of the assumed constitutive relations is investigated deriving the restrictions imposed by the second law of thermodynamics; remarkably, in the one--dimensional case, these constraints are explicitly solved.

In continuum thermodynamics, the entropy principle is of paramount importance  in the development of constitutive theories.
A systematic procedure for the exploitation of the second law was first developed in 1963 by Coleman and Noll \cite{ColNol}, where the entropy inequality was assumed in the form of the classical 
Clausius-Duhem inequality, the entropy flux being equal to the ratio between the  heat
flux and the absolute temperature.
Nevertheless, in the case where non-local constitutive relations are assumed, as we did in this paper, we need to extend Coleman-Noll procedure \cite{CimSelTri} by using as constraints both the field equations and some of their gradient extensions. Remarkably, this method  often provides an entropy extra-flux.

The plan of the paper is the  following. In Section~\ref{2}, the basic balance equations and the entropy inequality are introduced, the latter expressing, locally,
the second law of thermodynamics. In Section~\ref{3}, the extended Coleman-Noll approach is used to exploit the entropy principle, and the restrictions,
ensuring that the second law of thermodynamics is satisfied for arbitrary thermodynamical processes, are recovered \cite{CimSelTri}. In Section~\ref{4}, by considering
the one-dimensional case, a solution to the conditions on the constitutive quantities is provided. 
All the calculations, straightforward though long and tedious, which are necessary to derive and solve the constraints on the constitutive relations implied by the
entropy principle, are managed through the use of the 
Computer Algebra System Reduce \cite{Reduce}.
Accordingly, it is proved that the second law of thermodynamics permits the
constitutive equations to depend on all gradients that are included in the state space, and therefore  exhibit their compatibility with a general form of the Cauchy stress tensors of the RBCs and plasma components. Finally, Section~\ref{5} contains  some concluding remarks as well as possible future developments of the present approach.

\section{Mathematical formulation}\label{2}
\subsection{Field equations}
Let us suppose the blood to be made of a mixture consisting of red blood cells suspended in the plasma, so neglecting platelets, white blood cells (WBCs) and proteins.
Moreover, biochemical effects or mass interconversion are not
considered in this model. The volume fraction (or the concentration) of the RBCs is included in the field variables.

A suspension \cite{FranPalRog} is a molecular mixture where macromolecules 
replace solute molecules, thus allowing the treatment of suspensions as a kind of a limiting
case within the framework of thermodynamics of solid--fluid mixtures. In this context,  the total mass density of the two components mixture can be expressed as
\begin{equation}
\rho=\rho^{(1)}+\rho^{(2)}.
\end{equation}
This equation gives the commonly used expression for the mass density of a suspension, say
\begin{equation}
\rho=\phi\rho_p+(1-\phi)\rho_f,
\end{equation}
where
\begin{equation}\nonumber
\rho^{(1)}=\phi\rho_p, \qquad \rho^{(2)}=(1-\phi)\rho_f.
\end{equation}
Here, $\rho_p= \dfrac{\rho^{(1)}}{\phi}$ and
 $\rho_f = \dfrac{\rho^{(2)}}{1-\phi}$ are the pure densities of RBCs (solid particles) and  plasma (fluid component) in the reference configuration, respectively. The variable $\phi$  $(0\leq\phi<1)$ stands for the volume fraction of RBCs, that is the space-dependent concentration of RBCs  related to the hematocrit.
Trivially, the value $\phi=0$ is a limiting case corresponding to pure plasma. 
In the absence of chemical interactions, electromagnetic effects,
 and external forces, the field equations include  the balance of mass, linear momentum and energy.
These equations, with solid ($A=1$) and fluid ($A=2$) components, based on the
mixture theory \cite{True,Bow}, read
\begin{equation}
\label{eq-fields}
\begin{aligned}
&\rho^{(A)}_{,t}+\mathbf v^{(A)}\cdot\nabla\rho^{(A)}+\rho^{(A)} \nabla\cdot\mathbf v^{(A)}=0,\\ 
&\rho^{(A)}\left(\mathbf v^{(A)}_{,t}+\mathbf v^{(A)}\cdot \nabla\mathbf v^{(A)}\right)-\nabla\cdot \mathbf T^{(A)}=\mathbf{0},\\ 
&\rho^{(A)}\left(\varepsilon^{(A)}_{,t}+\mathbf v^{(A)}\cdot\nabla\varepsilon^{(A)}\right)-\mathbf T^{(A)}\cdot \nabla \mathbf v^{(A)}+\nabla\cdot \mathbf q^{(A)}=\rho^{(A)} r^{(A)},
\end{aligned}
\end{equation}
where the upscript ${}^{(A)}$ labels the two constituents, the subscript
${}_{,t}$ denotes the partial time derivative, 
$\mathbf v^{(A)}$ the velocities, $\varepsilon^{(A)}$ the partial internal energies per unit volume, $\mathbf T^{(A)}$ the partial Cauchy stress tensors, $\mathbf q^{(A)}$ the partial heat flux vectors, and $r^{(A)}$ the external heat supply of solid and fluid particles, respectively.
Moreover, in \eqref{eq-fields}, the interaction terms between the two components in the partial balances of linear momentum and internal energy are neglected.

Without couple stress tensors, the balance of angular momentum implies the 
symmetry of total Cauchy stress tensor; anyway, the partial stress tensors in general 
could not be symmetric.

In general, the constitutive relations
for $\mathbf T^{(1)}$ and $ \mathbf T^{(2)}$   are functions of the kinematical quantities associated to both constituents.
However, assuming  that the principle of phase separation applies, $\mathbf T^{(1)}$
and $\mathbf T^{(2)}$ can be taken dependent only on the kinematical quantities associated to red blood cells (for $A=1$) and plasma (for $A= 2$) (see \cite{Adkin, Adkin1}).
After assigning the partial stress tensors, a stress tensor for the suspension as a whole can be expressed as 
\begin{equation}
\mathbf T=\mathbf T^{(1)}+\mathbf T^{(2)},
\end{equation}
where 
\[
\mathbf T^{(1)}=\phi\mathbf{T}_p,\qquad \mathbf T^{(2)}=(1-\phi)\mathbf{T}_f,
\]
$\mathbf{T}_p$ and $\mathbf{T}_f$ being  the pure Cauchy stress tensors of solid particles (RBCs) and fluid component (plasma), respectively.
In addition, to account for additional dissipative effects,
we introduce two independent scalar
internal variables, indicated by $ \gamma^{(A)}$ ($A=1,2$),
whose time  evolution is  governed by equations of the form
\begin{equation}
\label{eq-gamma}
\rho^{(A)}\left(\frac{\partial\gamma^{(A)}}{\partial t}
+\mathbf{v}^{(A)}\cdot\nabla{\gamma^{(A)}}\right)=\Gamma^{(A)},\qquad A=1,2,
\end{equation}
where $\Gamma^{(A)}$ are suitable smooth functions.
Internal variables may represent an additional internal
degree of freedom of each phase, for instance, assuming the role of an
appropriate scalar microstructure  or another extensive feature \cite{Verhas}. 

\subsection{Entropy inequality} 
The complete analysis of the thermomechanical model requires, besides the consideration of the second law of thermodynamics,  some general invariance principles of the constitutive relations. The entropy inequality (in the Clausius-Duhem form) states that  the entropy
production has to be non-negative in every admissible process \cite{True-1957, Mul, CimJouRugVan, JCL}. The widely accepted interpretation is that the second law of
thermodynamics imposes restrictions on the type of motion and processes that are allowed, thus restricting the constitutive variables. Hence,  the constitutive equations must satisfy the entropy inequality for all solutions of the field equations.

Here, we shall assume the Clausius-Duhem to hold locally  for the whole blood.

Following the approach proposed in \cite{RajTao}, and assuming the temperatures of the  
 red blood cells  and  plasma to be different, let us consider the
 entropy productions for each constituent, based on classical irreversible thermodynamics \cite{Jou-bis}. 

Let  $s^{(A)}$, $\mathbf J_s^{(A)}$ and $\theta^{(A)}$ $(A=1,2)$ be
the partial specific entropies, the  entropy fluxes and  the temperatures of each constituent, respectively; 
the entropy productions  assume the following form
\begin{equation}
\sigma^{(A)}\equiv\rho^{(A)}\left(\frac{\partial s^{(A)}}{\partial t}+ \mathbf{v}^{(A)}\cdot\nabla s^{(A)}\right)+\nabla\cdot \mathbf J_s^{(A)}-\frac{1}{\theta^{(A)}}\rho^{(A)}r^{(A)},
\end{equation}  
with $A=1,2$. In general,  the components of the suspension may thermally interact in such a way the temperature or the velocity gradient
of one constituent could be sufficiently high to produce a negative entropy production for the other constituent.
Anyway, the entropy production of the blood suspension as a whole should remain  
non-negative for arbitrary thermodynamical processes, \emph{i.e.}, it is required that \cite{KirMas,BowGar}
\begin{equation}
\sigma=\sum_{A} \sigma^{(A)}\geq 0.
\end{equation}

\section{Thermodynamical restrictions}\label{3}
In this Section, we explore the consequences of the entropy principle on  the constitutive quantities, and prove that the restrictions provided by
applying the extended Coleman-Noll procedure \cite{CimSelTri} can be successfully
solved. Indeed, limiting ourselves to one-dimensional flows,
we proceed to the integration of the conditions placed by the
entropy inequality, and an explicit physically meaningful solution providing a representation of the constitutive equations is determined.

By virtue  of the axiom of material frame indifference, the material velocity  $\mathbf{v}^{(A)}$, and the anti-symmetric part of the velocity gradient $\mathbf W^{(A)} = (\nabla\mathbf{v}^{(A)})_{skew}$ cannot enter the set of state variables. Thence, {since we want to consider a first order non-local constitutive theory}, we assume the following state space
\begin{equation}
\label{statespace}
\mathcal Z\equiv \bigcup_{A=1}^2 \left\{\rho^{(A)},\theta^{(A)}, \gamma^{(A)},\mathbf{D}^{(A)},\nabla\rho^{(A)},\nabla\theta^{(A)},\nabla\gamma^{(A)}\right\},
\end{equation}
where $\mathbf{D}^{(A)} = (\nabla\mathbf{v}^{(A)})_{sym}$.
Consequently, the constitutive assumptions will be expressed as
\begin{equation*}
\Xi=\Xi(\mathbf z),\qquad \mathbf{z}\in\mathcal{Z},
\end{equation*}
where $\Xi$ represents any function in the set $\{ s^{(A)}, \mathbf{q}^{(A)}, r^{(A)}, \mathbf J_s^{(A)}, 
\mathbf{T}^{(A)},\Gamma^{(A)}\}$, with $A=1, 2$. 

In one space dimension and for each blood constituent, the balance equations of mass, linear momentum, energy and internal variable, given by equations \eqref{eq-fields} and \eqref{eq-gamma}, write in the form
\begin{align}
&\rho^{(1)}_{,t}+v^{(1)}\rho^{(1)}_{,x}+\rho^{(1)} v^{(1)}_{,x}=0,\nonumber\allowdisplaybreaks\\
&\rho^{(2)}_{,t}+v^{(2)}\rho^{(2)}_{,x}+\rho^{(2)} v^{(2)}_{,x}=0,\nonumber\allowdisplaybreaks\\
&\rho^{(1)}\left(v^{(1)}_{,t}+v^{(1)}v^{(1)}_{,x}\right)-T^{(1)}_{,x}=0,\nonumber\allowdisplaybreaks\\
&\rho^{(2)}\left(v^{(2)}_{,t}+v^{(2)}v^{(2)}_{,x}\right)-T^{(2)}_{,x}=0,\label{balance}\allowdisplaybreaks\\
&\rho^{(1)}\left(\varepsilon^{(1)}_{,t}+v^{(1)}\varepsilon^{(1)}_{,x}\right)-T^{(1)}v^{(1)}_{,x}+q^{(1)}_{,x}=0,\nonumber\allowdisplaybreaks\\
&\rho^{(2)}\left(\varepsilon^{(2)}_{,t}+v^{(2)}\varepsilon^{(2)}_{,x}\right)-T^{(2)}v^{(2)}_{,x}+q^{(2)}_{,x}=0,\nonumber\allowdisplaybreaks\\
&\rho^{(1)}\left(\gamma^{(1)}_{,t}+v^{(1)}\gamma^{(1)}_{,x}\right)=\Gamma^{(1)},\nonumber\allowdisplaybreaks\\
&\rho^{(2)}\left(\gamma^{(2)}_{,t}+v^{(2)}\gamma^{(2)}_{,x}\right)=\Gamma^{(2)},\nonumber
\end{align}
where external heat supplies are neglected.

The balance equations (\ref{balance}) must be closed assigning constitutive equations for
the partial Cauchy stress tensors $\mathbf{T}^{(A)}$, the partial heat fluxes $\mathbf{q}^{(A)}$, and the functions $\Gamma^{(A)}$  in such a way the local entropy
production of the whole blood,
\begin{equation}
\label{entropyinequality}
\sigma=\sigma^{(1)}+\sigma^{(2)}=\rho^{(1)} \left(s^{(1)}_{,t}+v^{(1)} s^{(1)}_{,x}\right)+ \rho^{(2)}\left(s^{(2)}_{,t}+v^{(2)} s^{(2)}_{,x}\right)+J^{(1)}_{s,x}+J^{(2)}_{s,x},
\end{equation}
be non-negative along any admissible thermodynamic process. {Of course, partial specific entropies and partial entropy fluxes are constitutive functions too.}

At the equilibrium, the  absolute temperatures of the RBCs and plasma constituents can be introduced, \emph{i.e.},
\begin{equation}
\label{absolute-temp}
\frac{1}{\theta^{(A)}}=\frac{\partial s^{(A)}}{\partial\varepsilon^{(A)}},\qquad A=1,2.
\end{equation}

Let us now define the Helmholtz free energies of the two constituents
\begin{equation}
\label{Helmholtz}
\psi^{(A)}=\varepsilon^{(A)}-\theta^{(A)}s^{(A)},\qquad
A=1,2,
\end{equation}
where $\psi^{(A)}$ is a constitutive quantity, and assume that $\varepsilon^{(A)}=\varepsilon^{(A)}(\rho^{(A)},\theta^{(A)})$.

On the basis of the chosen state space \eqref{statespace}, 
using relation (\ref{Helmholtz}) and expanding the derivatives by the chain rule, the entropy inequality (\ref{entropyinequality}) reads
\begin{align}
\label{entropyinequality-helmholtz}
&\left(\left(\frac{\partial\varepsilon^{(1)}}{\partial\rho^{(1)}}-\frac{\partial\psi^{(1)}}{\partial\rho^{(1)}}\right)\rho^{(1)}\theta^{(2)}-\frac{\partial\psi^{(2)}}{\partial\rho^{(1)}}\rho^{(2)}\theta^{(1)}\right)\rho^{(1)}_{,t}\notag\allowdisplaybreaks\\
+&\left(\left(\frac{\partial\varepsilon^{(2)}}{\partial\rho^{(2)}}-\frac{\partial\psi^{(2)}}{\partial\rho^{(2)}}\right)\rho^{(2)}\theta^{(1)}-\frac{\partial\psi^{(1)}}{\partial\rho^{(2)}}\rho^{(1)}\theta^{(2)}\right)\rho^{(2)}_{,t}\notag\allowdisplaybreaks\\
+&\left(\left(\frac{\partial\varepsilon^{(1)}}{\partial\theta^{(1)}}-\frac{\partial\psi^{(1)}}{\partial\theta^{(1)}}-\frac{\varepsilon^{(1)}-\psi^{(1)}}{\theta^{(1)}}\right)\rho^{(1)}\theta^{(2)}-\frac{\partial\psi^{(2)}}{\partial\theta^{(1)}}\rho^{(2)}\theta^{(1)}\right)\theta^{(1)}_{,t}\notag\allowdisplaybreaks\\
+&\left(\left(\frac{\partial\varepsilon^{(2)}}{\partial\theta^{(2)}}-\frac{\partial\psi^{(2)}}{\partial\theta^{(2)}}-\frac{\varepsilon^{(2)}-\psi^{(2)}}{\theta^{(2)}}\right)\rho^{(2)}\theta^{(1)}-\frac{\partial\psi^{(1)}}{\partial\theta^{(2)}}\rho^{(1)}\theta^{(2)}\right)\theta^{(2)}_{,t}\notag\allowdisplaybreaks\\
-&\left(\frac{\partial \psi^{(1)}}{\partial\gamma^{(1)}}\rho^{(1)}\theta^{(2)}+\frac{\partial \psi^{(2)}}{\partial\gamma^{(1)}}\rho^{(2)}\theta^{(1)}\right)\gamma^{(1)}_{,t}-\left(\frac{\partial \psi^{(1)}}{\partial\gamma^{(2)}}\rho^{(1)}\theta^{(2)}+\frac{\partial \psi^{(2)}}{\partial\gamma^{(2)}}\rho^{(2)}\theta^{(1)}\right)\gamma^{(2)}_{,t}\notag\allowdisplaybreaks\\
-&\left(\frac{\partial \psi^{(1)}}{\partial\rho^{(1)}_{,x}}\rho^{(1)}\theta^{(2)}+\frac{\partial \psi^{(2)}}{\partial\rho^{(1)}_{,x}}\rho^{(2)}\theta^{(1)}\right)\rho^{(1)}_{,tx}-\left(\frac{\partial \psi^{(1)}}{\partial\rho^{(2)}_{,x}}\rho^{(1)}\theta^{(2)}+\frac{\partial \psi^{(2)}}{\partial\rho^{(2)}_{,x}}\rho^{(2)}\theta^{(1)}\right)\rho^{(2)}_{,tx}\notag\allowdisplaybreaks\\
-&\left(\frac{\partial \psi^{(1)}}{\partial v^{(1)}_{,x}}\rho^{(1)}\theta^{(2)}+\frac{\partial \psi^{(2)}}{\partial v^{(1)}_{,x}}\rho^{(2)}\theta^{(1)}\right)v^{(1)}_{,tx}-\left(\frac{\partial \psi^{(1)}}{\partial v^{(2)}_{,x}}\rho^{(1)}\theta^{(2)}+\frac{\partial \psi^{(2)}}{\partial v^{(2)}_{,x}}\rho^{(2)}\theta^{(1)}\right) v^{(2)}_{,tx}\notag\allowdisplaybreaks\\
-&\left(\frac{\partial \psi^{(1)}}{\partial \theta^{(1)}_{,x}}\rho^{(1)}\theta^{(2)}+\frac{\partial \psi^{(2)}}{\partial \theta^{(1)}_{,x}}\rho^{(2)}\theta^{(1)}\right)\theta^{(1)}_{,tx}-\left(\frac{\partial \psi^{(1)}}{\partial \theta^{(2)}_{,x}}\rho^{(1)}\theta^{(2)}+\frac{\partial \psi^{(2)}}{\partial \theta^{(2)}_{,x}}\rho^{(2)}\theta^{(1)}\right) \theta^{(2)}_{,tx}\notag\allowdisplaybreaks\\
-&\left(\frac{\partial \psi^{(1)}}{\partial \gamma^{(1)}_{,x}}\rho^{(1)}\theta^{(2)}+\frac{\partial \psi^{(2)}}{\partial \gamma^{(1)}_{,x}}\rho^{(2)}\theta^{(1)}\right)\gamma^{(1)}_{,tx}-\left(\frac{\partial \psi^{(1)}}{\partial \gamma^{(2)}_{,x}}\rho^{(1)}\theta^{(2)}+\frac{\partial \psi^{(2)}}{\partial \gamma^{(2)}_{,x}}\rho^{(2)}\theta^{(1)}\right) \gamma^{(2)}_{,tx}\notag\allowdisplaybreaks\\
+&\left(\left(\frac{\partial J^{(1)}_s}{\partial \rho^{(1)}_{,x}}+\frac{\partial J^{(2)}_s}{\partial \rho^{(1)}_{,x}}\right)\theta^{(1)}\theta^{(2)}-\left(\frac{\partial \psi^{(1)}}{\partial \rho^{(1)}_{,x}}\rho^{(1)}v^{(1)}\theta^{(2)}+\frac{\partial \psi^{(2)}}{\partial \rho^{(1)}_{,x}}\rho^{(2)}v^{(2)}\theta^{(1)}\right)\right)\rho^{(1)}_{,xx}\notag\allowdisplaybreaks\\
+&\left(\left(\frac{\partial J^{(1)}_s}{\partial \rho^{(2)}_{,x}}+\frac{\partial J^{(2)}_s}{\partial \rho^{(2)}_{,x}}\right)\theta^{(1)}\theta^{(2)}-\left(\frac{\partial \psi^{(1)}}{\partial \rho^{(2)}_{,x}}\rho^{(1)}v^{(1)}\theta^{(2)}+\frac{\partial \psi^{(2)}}{\partial \rho^{(2)}_{,x}}\rho^{(2)}v^{(2)}\theta^{(1)}\right)\right)\rho^{(2)}_{,xx}\notag\allowdisplaybreaks\\
+&\left(\left(\frac{\partial J^{(1)}_s}{\partial v^{(1)}_{,x}}+\frac{\partial J^{(2)}_s}{\partial v^{(1)}_{,x}}\right)\theta^{(1)}\theta^{(2)}-\left(\frac{\partial \psi^{(1)}}{\partial v^{(1)}_{,x}}\rho^{(1)}v^{(1)}\theta^{(2)}+\frac{\partial \psi^{(2)}}{\partial v^{(1)}_{,x}}\rho^{(2)}v^{(2)}\theta^{(1)}\right)\right)v^{(1)}_{,xx}\notag\allowdisplaybreaks\\
+&\left(\left(\frac{\partial J^{(1)}_s}{\partial v^{(2)}_{,x}}+\frac{\partial J^{(2)}_s}{\partial v^{(2)}_{,x}}\right)\theta^{(1)}\theta^{(2)}-\left(\frac{\partial \psi^{(1)}}{\partial v^{(2)}_{,x}}\rho^{(1)}v^{(1)}\theta^{(2)}+\frac{\partial \psi^{(2)}}{\partial v^{(2)}_{,x}}\rho^{(2)}v^{(2)}\theta^{(1)}\right)\right)v^{(2)}_{,xx}\notag\allowdisplaybreaks\\
+&\left(\left(\frac{\partial J^{(1)}_s}{\partial \theta^{(1)}_{,x}}+\frac{\partial J^{(2)}_s}{\partial \theta^{(1)}_{,x}}\right)\theta^{(1)}\theta^{(2)}-\left(\frac{\partial \psi^{(1)}}{\partial \theta^{(1)}_{,x}}\rho^{(1)}v^{(1)}\theta^{(2)}+\frac{\partial \psi^{(2)}}{\partial \theta^{(1)}_{,x}}\rho^{(2)}v^{(2)}\theta^{(1)}\right)\right)\theta^{(1)}_{,xx}\notag\allowdisplaybreaks\\
+&\left(\left(\frac{\partial J^{(1)}_s}{\partial \theta^{(2)}_{,x}}+\frac{\partial J^{(2)}_s}{\partial \theta^{(2)}_{,x}}\right)\theta^{(1)}\theta^{(2)}-\left(\frac{\partial \psi^{(1)}}{\partial \theta^{(2)}_{,x}}\rho^{(1)}v^{(1)}\theta^{(2)}+\frac{\partial \psi^{(2)}}{\partial \theta^{(2)}_{,x}}\rho^{(2)}v^{(2)}\theta^{(1)}\right)\right)\theta^{(2)}_{,xx}\notag\allowdisplaybreaks\\
+&\left(\left(\frac{\partial J^{(1)}_s}{\partial \gamma^{(1)}_{,x}}+\frac{\partial J^{(2)}_s}{\partial \gamma^{(1)}_{,x}}\right)\theta^{(1)}\theta^{(2)}-\left(\frac{\partial \psi^{(1)}}{\partial \gamma^{(1)}_{,x}}\rho^{(1)}v^{(1)}\theta^{(2)}+\frac{\partial \psi^{(2)}}{\partial \gamma^{(1)}_{,x}}\rho^{(2)}v^{(2)}\theta^{(1)}\right)\right)\gamma^{(1)}_{,xx}\notag\allowdisplaybreaks\\
+&\left(\left(\frac{\partial J^{(1)}_s}{\partial \gamma^{(2)}_{,x}}+\frac{\partial J^{(2)}_s}{\partial \gamma^{(2)}_{,x}}\right)\theta^{(1)}\theta^{(2)}-\left(\frac{\partial \psi^{(1)}}{\partial \gamma^{(2)}_{,x}}\rho^{(1)}v^{(1)}\theta^{(2)}+\frac{\partial \psi^{(2)}}{\partial \gamma^{(2)}_{,x}}\rho^{(2)}v^{(2)}\theta^{(1)}\right)\right)\gamma^{(2)}_{,xx}\notag\allowdisplaybreaks\\
+&\left(\left(\frac{\partial J^{(1)}_s}{\partial \rho^{(1)}}+\frac{\partial J^{(2)}_s}{\partial \rho^{(1)}}\right)\theta^{(1)}\theta^{(2)}+\frac{\partial\varepsilon^{(1)}}{\partial\rho^{(1)}}\rho^{(1)}v^{(1)}\theta^{(2)}\notag\right.\\
-&\left.\left(\frac{\partial \psi^{(1)}}{\partial \rho^{(1)}}\rho^{(1)}v^{(1)}\theta^{(2)}+\frac{\partial \psi^{(2)}}{\partial \rho^{(1)}}\rho^{(2)}v^{(2)}\theta^{(1)}\right)\right)\rho^{(1)}_{,x}\notag\allowdisplaybreaks\\
+&\left(\left(\frac{\partial J^{(1)}_s}{\partial \rho^{(2)}}+\frac{\partial J^{(2)}_s}{\partial \rho^{(2)}}\right)\theta^{(1)}\theta^{(2)}+\frac{\partial\varepsilon^{(2)}}{\partial\rho^{(2)}}\rho^{(2)}v^{(2)}\theta^{(1)}\notag\right.\\
-&\left.\left(\frac{\partial \psi^{(1)}}{\partial \rho^{(2)}}\rho^{(1)}v^{(1)}\theta^{(2)}+\frac{\partial \psi^{(2)}}{\partial \rho^{(2)}}\rho^{(2)}v^{(2)}\theta^{(1)}\right)\right)\rho^{(2)}_{,x}\notag\allowdisplaybreaks\\
+&\left(\left(\frac{\partial J^{(1)}_s}{\partial \theta^{(1)}}+\frac{\partial J^{(2)}_s}{\partial \theta^{(1)}}\right)\theta^{(1)}\theta^{(2)}+\frac{\partial\varepsilon^{(1)}}{\partial\theta^{(1)}}\rho^{(1)}v^{(1)}\theta^{(2)}\notag\right.\allowdisplaybreaks\\
-&\left.\left(\frac{\partial \psi^{(1)}}{\partial \theta^{(1)}}\rho^{(1)}v^{(1)}\theta^{(2)}+\frac{\partial \psi^{(2)}}{\partial \theta^{(1)}}\rho^{(2)}v^{(2)}\theta^{(1)}\right)-\frac{\varepsilon^{(1)}-\psi^{(1)}}{\theta^{(1)}}\rho^{(1)}v^{(1)}\theta^{(2)}\right)\theta^{(1)}_{,x}\notag\allowdisplaybreaks\\
+&\left(\left(\frac{\partial J^{(1)}_s}{\partial \theta^{(2)}}+\frac{\partial J^{(2)}_s}{\partial \theta^{(2)}}\right)\theta^{(1)}\theta^{(2)}+\frac{\partial\varepsilon^{(2)}}{\partial\theta^{(2)}}\rho^{(2)}v^{(2)}\theta^{(1)}\notag\right.\allowdisplaybreaks\\
-&\left.\left(\frac{\partial \psi^{(1)}}{\partial \theta^{(2)}}\rho^{(1)}v^{(1)}\theta^{(2)}+\frac{\partial \psi^{(2)}}{\partial \theta^{(2)}}\rho^{(2)}v^{(2)}\theta^{(1)}\right)-\frac{\varepsilon^{(2)}-\psi^{(2)}}{\theta^{(2)}}\rho^{(2)}v^{(2)}\theta^{(1)}\right)\theta^{(2)}_{,x}\notag\allowdisplaybreaks\\
+&\left(\left(\frac{\partial J^{(1)}_s}{\partial \gamma^{(1)}}+\frac{\partial J^{(2)}_s}{\partial \gamma^{(1)}}\right)\theta^{(1)}\theta^{(2)}-\left(\frac{\partial \psi^{(1)}}{\partial \gamma^{(1)}}\rho^{(1)}v^{(1)}\theta^{(2)}+\frac{\partial \psi^{(2)}}{\partial \gamma^{(1)}}\rho^{(2)}v^{(2)}\theta^{(1)}\right)\right)\gamma^{(1)}_{,x}\notag\allowdisplaybreaks\\
+&\left(\left(\frac{\partial J^{(1)}_s}{\partial \gamma^{(2)}}+\frac{\partial J^{(2)}_s}{\partial \gamma^{(2)}}\right)\theta^{(1)}\theta^{(2)}-\left(\frac{\partial \psi^{(1)}}{\partial \gamma^{(2)}}\rho^{(1)}v^{(1)}\theta^{(2)}+\frac{\partial \psi^{(2)}}{\partial \gamma^{(2)}}\rho^{(2)}v^{(2)}\theta^{(1)}\right)\right)\gamma^{(2)}_{,x}\geq 0.
\end{align}
The latter inequality has to be satisfied for all admissible thermodynamical processes, which in turn must be solutions of equations \eqref{balance}; thus, time derivatives of the field variables need to be eliminated using 
\eqref{balance}. Moreover, since also time derivatives of first order space derivatives of some field variables are involved in \eqref{entropyinequality-helmholtz}, then we need to eliminate them using some differential consequences of the field equations. In fact, among all admissible processes, there are smooth processes and these must satisfy both the field equations and their differential consequences. These considerations justify from a mathematical point of view the extended Coleman-Noll procedure \cite{CimSelTri}, where we need to use as constraints in the entropy inequality \eqref{entropyinequality-helmholtz} both the balance equations \eqref{balance} and some of their gradient extensions; in our case, because the state space contain first order gradients, it is sufficient to consider  the first order gradients of balance equations.

Thus, eliminating all time derivatives in \eqref{entropyinequality-helmholtz},  we get an entropy inequality where we can recognize the so-called \emph{highest} derivatives, say 
\begin{equation}
\mathbf{X}=\left\{\rho^{(1)}_{,xxx},\rho^{(2)}_{,xxx},v^{(1)}_{,xxx},v^{(2)}_{,xxx},\theta^{(1)}_{,xxx},\theta^{(2)}_{,xxx},\gamma^{(1)}_{,xxx},\gamma^{(2)}_{,xxx}\right\},
\end{equation}
and the \emph{higher} ones, namely
\begin{equation}
\mathbf{Y}=\left\{\rho^{(1)}_{,xx},\rho^{(2)}_{,xx},v^{(1)}_{,xx},v^{(2)}_{,xx},\theta^{(1)}_{,xx},\theta^{(2)}_{,xx},\gamma^{(1)}_{,xx},\gamma^{(2)}_{,xx}\right\}.
\end{equation}
Highest derivatives are the spatial derivatives of maximal order, whereas higher derivatives are spatial derivatives, not of maximal order, whose order is higher than that entering the variables of the state space. Remarkably, they may assume arbitrary values \cite{CimSelTri}.

After straightforward though lengthy computations, the inequality \eqref{entropyinequality-helmholtz} can be expressed in the following compact form
\begin{equation}
\label{entropycompatta}
\sum_p A_p X_p+\sum_{q,r}B_{qr}Y_qY_r+\sum_q C_qY_q+D\ge 0,
\end{equation}
where the quantities $A_p$ and $C_{q}$ are the components of vectors having the dimension of highest and higher derivatives, respectively, $B_{qr}$ are the entries of a matrix of order equal to the number of higher derivatives, and $D$ is a scalar function; all these quantities depend only on the field and state space variables.
Thus, entropy inequality is linear in the highest derivatives and quadratic in the higher ones \cite{GorOl1Rog}.
The values of the highest and higher derivatives can be arbitrarily attributed regardless of the value of $D$, which, rather, is defined on the state space. 
First of all, let us observe that in principle nothing forbids the possibility of 
a thermodynamic process in which $D=0$ \cite{CimOlTri}. Thus, the inequality \eqref{entropycompatta} has to be satisfied 
for arbitrary $X_{p}$ and $Y_q$, and the conditions 
\begin{equation}
A_p=0, \qquad C_{q}=0,\qquad D\ge 0,
\end{equation} 
as well as the constraint that the quantities $B_{qr}$ are the entries of a positive semidefinite symmetric matrix, are necessary and sufficient for the achievement of the entropy inequality (see \cite{CimOlTri}).

By annihilating the coefficients of the highest derivatives, \emph{i.e.}, $A_p=0$, we get the restrictions
\begin{align}
&\rho^{(2)}\frac{\partial \varepsilon^{(2)}}{\partial \theta^{(2)}}\left(\rho^{(1)}\theta^{(2)}\frac{\partial \psi^{(1)}}{\partial\theta^{(1)}_{,x}}+\rho^{(2)}\theta^{(1)}\frac{\partial \psi^{(2)}}{\partial\theta^{(1)}_{,x}}\right)\frac{\partial q^{(1)}}{\partial\rho^{(A)}_{,x}}\notag\allowdisplaybreaks\\
&+\rho^{(1)}\frac{\partial \varepsilon^{(1)}}{\partial \theta^{(1)}}\left(\rho^{(1)}\theta^{(2)}\frac{\partial \psi^{(1)}}{\partial\theta^{(2)}_{,x}}+\rho^{(2)}\theta^{(1)}\frac{\partial \psi^{(2)}}{\partial\theta^{(2)}_{,x}}\right)\frac{\partial q^{(2)}}{\partial\rho^{(A)}_{,x}}\nonumber\allowdisplaybreaks\\
&-\rho^{(2)}\frac{\partial \varepsilon^{(1)}}{\partial \theta^{(1)}}\frac{\partial \varepsilon^{(2)}}{\partial \theta^{(2)}}\left(\rho^{(1)}\theta^{(2)}\frac{\partial \psi^{(1)}}{\partial v^{(1)}_{,x}}+\rho^{(2)}\theta^{(1)}\frac{\partial \psi^{(2)}}{\partial v^{(1)}_{,x}}\right)\frac{\partial T^{(1)}}{\partial\rho^{(A)}_{,x}}\nonumber\allowdisplaybreaks\\
&-\rho^{(1)}\frac{\partial \varepsilon^{(1)}}{\partial \theta^{(1)}}\frac{\partial \varepsilon^{(2)}}{\partial \theta^{(2)}}\left(\rho^{(1)}\theta^{(2)}\frac{\partial \psi^{(1)}}{\partial v^{(2)}_{,x}}+\rho^{(2)}\theta^{(1)}\frac{\partial \psi^{(2)}}{\partial v^{(2)}_{,x}}\right)\frac{\partial T^{(2)}}{\partial\rho^{(A)}_{,x}}=0,\allowdisplaybreaks \label{highest-cond-1}\\
&\rho^{(2)}\frac{\partial \varepsilon^{(2)}}{\partial \theta^{(2)}}\left(\rho^{(1)}\theta^{(2)}\frac{\partial \psi^{(1)}}{\partial\theta^{(1)}_{,x}}+\rho^{(2)}\theta^{(1)}\frac{\partial \psi^{(2)}}{\partial\theta^{(1)}_{,x}}\right)\frac{\partial q^{(1)}}{\partial v^{(A)}_{,x}}\nonumber\allowdisplaybreaks\\
&+\rho^{(1)}\frac{\partial \varepsilon^{(1)}}{\partial \theta^{(1)}}\left(\rho^{(1)}\theta^{(2)}\frac{\partial \psi^{(1)}}{\partial\theta^{(2)}_{,x}}+\rho^{(2)}\theta^{(1)}\frac{\partial \psi^{(2)}}{\partial\theta^{(2)}_{,x}}\right)\frac{\partial q^{(2)}}{\partial v^{(A)}_{,x}}\nonumber\allowdisplaybreaks\\
&-\rho^{(2)}\frac{\partial \varepsilon^{(1)}}{\partial \theta^{(1)}}\frac{\partial \varepsilon^{(2)}}{\partial \theta^{(2)}}\left(\rho^{(1)}\theta^{(2)}\frac{\partial \psi^{(1)}}{\partial v^{(1)}_{,x}}+\rho^{(2)}\theta^{(1)}\frac{\partial \psi^{(2)}}{\partial v^{(1)}_{,x}}\right)\frac{\partial T^{(1)}}{\partial v^{(A)}_{,x}}\nonumber\allowdisplaybreaks\\
&-\rho^{(1)}\frac{\partial \varepsilon^{(1)}}{\partial \theta^{(1)}}\frac{\partial \varepsilon^{(2)}}{\partial \theta^{(2)}}\left(\rho^{(1)}\theta^{(2)}\frac{\partial \psi^{(1)}}{\partial v^{(2)}_{,x}}+\rho^{(2)}\theta^{(1)}\frac{\partial \psi^{(2)}}{\partial v^{(2)}_{,x}}\right)\frac{\partial T^{(2)}}{\partial v^{(A)}_{,x}}=0,\allowdisplaybreaks\\
&\rho^{(2)}\frac{\partial \varepsilon^{(2)}}{\partial \theta^{(2)}}\left(\rho^{(1)}\theta^{(2)}\frac{\partial \psi^{(1)}}{\partial\theta^{(1)}_{,x}}+\rho^{(2)}\theta^{(1)}\frac{\partial \psi^{(2)}}{\partial\theta^{(1)}_{,x}}\right)\frac{\partial q^{(1)}}{\partial\theta^{(A)}_{,x}}\nonumber\allowdisplaybreaks\\
&+\rho^{(1)}\frac{\partial \varepsilon^{(1)}}{\partial \theta^{(1)}}\left(\rho^{(1)}\theta^{(2)}\frac{\partial \psi^{(1)}}{\partial\theta^{(2)}_{,x}}+\rho^{(2)}\theta^{(1)}\frac{\partial \psi^{(2)}}{\partial\theta^{(2)}_{,x}}\right)\frac{\partial q^{(2)}}{\partial\theta^{(A)}_{,x}}\nonumber\allowdisplaybreaks\\
&-\rho^{(2)}\frac{\partial \varepsilon^{(1)}}{\partial \theta^{(1)}}\frac{\partial \varepsilon^{(2)}}{\partial \theta^{(2)}}\left(\rho^{(1)}\theta^{(2)}\frac{\partial \psi^{(1)}}{\partial v^{(1)}_{,x}}+\rho^{(2)}\theta^{(1)}\frac{\partial \psi^{(2)}}{\partial v^{(1)}_{,x}}\right)\frac{\partial T^{(1)}}{\partial\theta^{(A)}_{,x}}\nonumber\allowdisplaybreaks\\
&-\rho^{(1)}\frac{\partial \varepsilon^{(1)}}{\partial \theta^{(1)}}\frac{\partial \varepsilon^{(2)}}{\partial \theta^{(2)}}\left(\rho^{(1)}\theta^{(2)}\frac{\partial \psi^{(1)}}{\partial v^{(2)}_{,x}}+\rho^{(2)}\theta^{(1)}\frac{\partial \psi^{(2)}}{\partial v^{(2)}_{,x}}\right)\frac{\partial T^{(2)}}{\partial\theta^{(A)}_{,x}}=0,\allowdisplaybreaks\\
&\rho^{(2)}\frac{\partial \varepsilon^{(2)}}{\partial \theta^{(2)}}\left(\rho^{(1)}\theta^{(2)}\frac{\partial \psi^{(1)}}{\partial\theta^{(1)}_{,x}}+\rho^{(2)}\theta^{(1)}\frac{\partial \psi^{(2)}}{\partial\theta^{(1)}_{,x}}\right)\frac{\partial q^{(1)}}{\partial\gamma^{(A)}_{,x}}\nonumber\allowdisplaybreaks\\
&+\rho^{(1)}\frac{\partial \varepsilon^{(1)}}{\partial \theta^{(1)}}\left(\rho^{(1)}\theta^{(2)}\frac{\partial \psi^{(1)}}{\partial\theta^{(2)}_{,x}}+\rho^{(2)}\theta^{(1)}\frac{\partial \psi^{(2)}}{\partial\theta^{(2)}_{,x}}\right)\frac{\partial q^{(2)}}{\partial\gamma^{(A)}_{,x}}\nonumber\allowdisplaybreaks\\
&-\rho^{(2)}\frac{\partial \varepsilon^{(1)}}{\partial \theta^{(1)}}\frac{\partial \varepsilon^{(2)}}{\partial \theta^{(2)}}\left(\rho^{(1)}\theta^{(2)}\frac{\partial \psi^{(1)}}{\partial v^{(1)}_{,x}}+\rho^{(2)}\theta^{(1)}\frac{\partial \psi^{(2)}}{\partial v^{(1)}_{,x}}\right)\frac{\partial T^{(1)}}{\partial\gamma^{(A)}_{,x}}\nonumber\allowdisplaybreaks\\
&-\rho^{(1)}\frac{\partial \varepsilon^{(1)}}{\partial \theta^{(1)}}\frac{\partial \varepsilon^{(2)}}{\partial \theta^{(2)}}\left(\rho^{(1)}\theta^{(2)}\frac{\partial \psi^{(1)}}{\partial v^{(2)}_{,x}}+\rho^{(2)}\theta^{(1)}\frac{\partial \psi^{(2)}}{\partial v^{(2)}_{,x}}\right)\frac{\partial T^{(2)}}{\partial\gamma^{(A)}_{,x}}=0,\label{highest-cond-4}
\end{align}
where $A=1,2$. The thermodynamical restrictions \eqref{highest-cond-1}-\eqref{highest-cond-4} are satisfied provided that
\begin{equation}
\label{cond-s1-s2}
\rho^{(1)}\theta^{(2)}\psi^{(1)}+\rho^{(2)}\theta^{(1)}\psi^{(2)}=F(\rho^{(1)},\rho^{(2)},\theta^{(1)},\theta^{(2)},\gamma^{(1)},\gamma^{(2)},\rho^{(1)}_{,x},\rho^{(2)}_{,x},\gamma^{(1)}_{,x},\gamma^{(2)}_{,x}),
\end{equation}
being $F$ an arbitrary function depending on its arguments.

Furthermore, annihilating the coefficients of the linear terms in the higher derivatives, \emph{i.e.}, $C_{q}=0$, further thermodynamic restrictions are obtained, involving the Cauchy stress tensors,  heat fluxes,  specific entropies, and entropy fluxes of both phases, $\Gamma^{(1)}$ and $\Gamma^{(2)}$; even if their derivation is straightforward, we omit to write here  their expressions which are  rather long. It is worth of being remarked that we obtain that the entropy fluxes  no longer assume  the classical  expression  postulated in rational thermodynamics, whereupon extra-flux contributions arise.

\section{A solution to the thermodynamic restrictions }\label{4}
The thermodynamical restrictions derived are still
too general to be useful in practical applications; therefore,  suitable simplifications are needed depending on the speciﬁc situations we
want to describe \cite{GorOlRog}. Then, in order to proceed with the exploitation of the
entropy inequality for the model of a solid-fluid mixture, let
us assume the following constitutive laws for $T^{(1)}$, $T^{(2)}$, $q^{(1)}$, $q^{(2)}$, $\Gamma^{(1)}$ and $\Gamma^{(2)}$:
\begin{equation}
\begin{aligned}
{T^{(1)}}&={\tau^{(1)}_0+\tau^{(1)}_1 (\rho^{(1)}_{,x})^2+\tau^{(1)}_2 \rho^{(1)}_{,x}\gamma^{(1)}_{,x}+\tau^{(1)}_3 (\gamma^{(1)}_{,x})^2+\tau^{(1)}_4v^{(1)}_{,x}+\tau^{(1)}_5v^{(2)}_{,x},}\\
{T^{(2)}}&={\tau^{(2)}_0+\tau^{(2)}_3 (\gamma^{(2)}_{,x})^2+\tau^{(2)}_4v^{(1)}_{,x}+\tau^{(2)}_5v^{(2)}_{,x},}\\
q^{(A)}&=q^{(A)}_1\varepsilon^{(1)}_{,x}+q^{(A)}_2\varepsilon^{(2)}_{,x}+q^{(A)}_3\rho^{(1)}_{,x}+q^{(A)}_4\rho^{(2)}_{,x}\\
&=q^{(A)}_1\left(\frac{\partial \varepsilon^{(1)}}{\partial \rho^{(1)}}\rho^{(1)}_{,x}+\frac{\partial \varepsilon^{(1)}}{\partial \theta^{(1)}}\theta^{(1)}_{,x}\right)+q^{(A)}_2\left(\frac{\partial \varepsilon^{(2)}}{\partial \rho^{(2)}}\rho^{(2)}_{,x}+\frac{\partial \varepsilon^{(2)}}{\partial \theta^{(2)}}\theta^{(2)}_{,x}\right)\\
&+q^{(A)}_3\rho^{(1)}_{,x}+q^{(A)}_4\rho^{(2)}_{,x},\\
\Gamma^{(A)}&=\widehat\Gamma^{(A)}_0+\widehat\Gamma^{(A)}_1 (\rho^{(A)}_{,x})^2+\widehat\Gamma^{(A)}_2 \rho^{(A)}_{,x}\gamma^{(A)}_{,x}+\widehat \Gamma^{(A)}_3 (\gamma^{(A)}_{,x})^2+\widehat\Gamma^{(A)}_4v^{(A)}_{,x},
\end{aligned}
\end{equation}
with $A=1,2$, where 
\begin{equation}
\begin{aligned}
&\tau^{(1)}_i\equiv\tau^{(1)}_i(\rho^{(1)},\rho^{(2)},\theta^{(1)},\theta^{(2)},\gamma^{(1)},\gamma^{(2)}),\qquad &&i=0,\ldots,5,\\
&\tau^{(2)}_j\equiv\tau^{(2)}_j(\rho^{(1)},\rho^{(2)},\theta^{(1)},\theta^{(2)},\gamma^{(1)},\gamma^{(2)}),\qquad &&j=0,3,4,5,\\
&q^{(A)}_k\equiv q^{(A)}_k(\rho^{(1)},\rho^{(2)},\theta^{(1)},\theta^{(2)},\gamma^{(1)},\gamma^{(2)}),\qquad &&k=1,\ldots,4,\\
&\widehat\Gamma^{(A)}_\ell\equiv \widehat\Gamma^{(A)}_\ell(\rho^{(1)},\rho^{(2)},\theta^{(1)},\theta^{(2)},\gamma^{(1)},\gamma^{(2)}),\qquad &&\ell=0,\ldots,4,
\end{aligned}
\end{equation}
are some smooth functions of their arguments.
Moreover, let us expand the partial Helmholtz free energies around the homogeneous equilibrium state (where all the gradients vanish) at the first order on some gradients of the field  variables entering the state space, chosen according to the constraint (\ref{cond-s1-s2}), \emph{i.e.},
\begin{equation}
\begin{aligned}
&\psi^{(A)}=\widehat{\psi}^{(A)}_0+\widehat{\psi}^{(A)}_1(\rho^{(A)}_{,x})^2+\widehat{\psi}^{(A)}_2\rho^{(A)}_{,x}\gamma^{(A)}_{,x}+\widehat{\psi}^{(A)}_3(\gamma^{(A)}_{,x})^2,
\end{aligned}
\end{equation}
where $\widehat{\psi}^{(A)}_k\equiv \widehat{\psi}^{(A)}_k(\rho^{(A)},\theta^{(A)},\gamma^{(A)})$, with $k=0,\ldots,3$, are some functions of their arguments.

On the basis of these assumptions, the thermodynamical restrictions { $C_q=0$, obtained by annihilating the coefficients of the higher derivatives,} yield a large set of partial differential equations that we manage using some routines written
in the CAS Reduce \cite{Reduce}; we are able to recover  the form of the coefficients
entering the partial Helmholtz free energies as well as the functions $\Gamma^{(A)}$, namely
\begin{equation}
\begin{aligned}
&\widehat{\psi}^{(A)}_0= \varepsilon^{(A)}-\theta^{(A)}\psi^{(A)}_0,\qquad&&\widehat{\psi}^{(A)}_1=-\theta^{(A)}\frac{\psi^{(A)}_3(\Gamma^{(A)}_4)^2}{(\rho^{(A)})^4},\\
&\widehat{\psi}^{(A)}_2=-2\theta^{(A)}\frac{\psi^{(A)}_3\Gamma^{(A)}_4}{(\rho^{(A)})^2},\qquad &&\widehat{\psi}^{(A)}_3= -\theta^{(A)}{\psi}^{(A)}_3,\\
&\widehat\Gamma^{(A)}_0= \Gamma^{(A)}_0, \qquad &&\widehat\Gamma^{(A)}_1=\kappa^{(A)}\frac{(\psi^{(A)}_3)^2(\Gamma^{(A)}_4)^2}{\rho^{(A)}},\\
&\widehat\Gamma^{(A)}_2=2\kappa^{(A)} \rho^{(A)} (\psi^{(A)}_3)^2\Gamma^{(A)}_4,\qquad &&\widehat\Gamma^{(A)}_4= \Gamma^{(A)}_4,
\end{aligned}
\end{equation}
where $\psi^{(A)}_0\equiv \psi^{(A)}_0(\rho^{(A)},\theta^{(A)})$, $\psi^{(A)}_3\equiv \psi^{(A)}_3(\rho^{(A)})$, 
$\Gamma^{(A)}_0\equiv \Gamma^{(A)}_0(\rho^{(A)},\theta^{(A)},\gamma^{(A)})$, $\Gamma^{(A)}_4\equiv \Gamma^{(A)}_4(\rho^{(A)})$, and $\kappa^{(A)}\,\, (A=1,2)$ are arbitrary constants.

Using the principle of maximum entropy at the
equilibrium, in the expression of the total entropy, 
\begin{equation}
\begin{aligned}
\rho^{(1)}{s}^{(1)}+\rho^{(2)}{s}^{(2)}&=\rho^{(1)}\psi^{(1)}_0+\rho^{(2)}\psi^{(2)}_0\\
&+\frac{\psi^{(1)}_3}{(\rho^{(1)})^3}\left(\Gamma^{(1)}_4\rho^{(1)}_{,x}+(\rho^{(1)})^2\gamma^{(1)}_{,x}\right)^2\\
&+\frac{\psi^{(2)}_3}{(\rho^{(2)})^3}\left(\Gamma^{(2)}_4\rho^{(2)}_{,x}+(\rho^{(2)})^2\gamma^{(2)}_{,x}\right)^2,
\end{aligned}
\end{equation}
it is required that the quadratic part in the gradients
must be negative semidefinite; whereupon, it has to be
\begin{equation}
\label{max-entropy}
\psi^{(A)}_3\le 0,\qquad A=1,2.
\end{equation}
Furthermore, we get also the representation of the partial entropy fluxes, say
\begin{equation}
\label{js}
\begin{aligned}
J^{(A)}_s&=\frac{q^{(A)}}{\theta^{(A)}}-\frac{4}{3}\kappa^{(A)}\frac{(\psi^{(A)}_3)^3}{(\rho^{(A)})^3}\left(\Gamma^{(A)}_4\rho^{(A)}_{,x}+(\rho^{(A)})^2\gamma^{(A)}_{,x}\right)^3,
\end{aligned}
\end{equation}
where, in addition to the classical term given by the ratio between the partial heat
flux and the partial absolute temperature, an entropy extra-flux  incorporating cubic terms in the first order gradients $\rho^{(1)}_{,x}$, $\rho^{(2)}_{,x}$, $\gamma^{(1)}_{,x}$ and $\gamma^{(2)}_{,x}$ can be recognized.

Using these results, the entries $B_{qr}$ turn out to be identically zero, and the entropy inequality reduces to
a cubic polynomial in the gradients $\rho^{(1)}_{,x}$, $\rho^{(2)}_{,x}$, $v^{(1)}_{,x}$, $v^{(2)}_{,x}$, $\theta^{(1)}_{,x}$, $\theta^{(2)}_{,x}$, $\gamma^{(1)}_{,x}$ and $\gamma^{(2)}_{,x}$.
Entropy principle is not violated provided that  linear and cubic terms in such gradients  vanish, whence the material functions entering the two partial Cauchy stress tensors can be written as follows:
\begin{equation}
\label{cauchy-comp}
\begin{aligned}
\tau^{(A)}_0&=(\rho^{(A)})^2\left(\theta^{(A)} \frac{\partial \psi^{(A)}_0}{\partial \rho^{(A)}}-\frac{\partial \varepsilon^{(A)}}{\partial \rho^{(A)}}\right),\qquad  A=1,2,\\
\tau^{(1)}_1&=\theta^{(1)}\frac{(\Gamma^{(1)}_4)^2}{(\rho^{(1)})^3}\left(\rho^{(1)}\frac{\partial \psi^{(1)}_3}{\partial \rho^{(1)}}+2\psi^{(1)}_3\right),\\
\tau^{(1)}_2&=2\theta^{(1)}\frac{\Gamma^{(1)}_4}{\rho^{(1)}}\left(\rho^{(1)}\frac{\partial \psi^{(1)}_3}{\partial \rho^{(1)}}+2\psi^{(1)}_3\right),\\
\tau^{(A)}_3&=\theta^{(A)}\rho^{(A)}\left(\rho^{(A)}\frac{\partial \psi^{(A)}_3}{\partial \rho^{(A)}}+2\psi^{(A)}_3\right),\qquad  A=1,2,
\end{aligned}
\end{equation}
and the following constraint is also obtained
\begin{equation}
\label{gamma2}
\Gamma^{(2)}_4=0.
\end{equation}
What remains of the entropy inequality is the following quadratic form in the gradients entering the state space, say 
\begin{equation}
\label{residual-last}
\begin{aligned}
&q^{(1)}\frac{\partial}{\partial x}\left(\frac{1}{\theta^{(1)}}\right)+q^{(2)}\frac{\partial}{\partial x}\left(\frac{1}{\theta^{(2)}}\right)+\frac{\tau^{(1)}_4}{\theta^{(1)}}(v^{(1)}_{,x})^2+\left(\frac{\tau^{(1)}_5}{\theta^{(1)}}+\frac{\tau^{(2)}_4}{\theta^{(2)}}\right)v^{(1)}_{,x}v^{(2)}_{,x}+\frac{\tau^{(2)}_5}{\theta^{(2)}}(v^{(2)}_{,x})^2\\
&+2\frac{\psi^{(1)}_3}{\rho^{(1)}}\left(\Gamma^{(1)}_4\rho^{(1)}_{,x}+(\rho^{(1)})^2\gamma^{(1)}_{,x}\right)\frac{\partial}{\partial x}\left(\frac{\Gamma^{(1)}_0}{\rho^{(1)}}\right)+2\psi^{(2)}_3\rho^{(2)}\gamma^{(2)}_{,x}\frac{\partial}{\partial x}\left(\frac{\Gamma^{(2)}_0}{\rho^{(2)}}\right)\geq 0,
\end{aligned}
\end{equation}
whose coefficients depend at most on the field variables,
provided that 
\begin{equation}
\label{gamma3}
\widehat\Gamma^{(A)}_3=\kappa^{(A)} (\rho^{(A)})^3 (\psi^{(A)}_3)^2.
\end{equation}
Constraint \eqref{gamma3} is not a thermodynamical restriction derived from the compatibility with the second law of thermodynamics; however, we assume it for technical reasons, since it allows the reduced entropy inequality to be a homogeneous quadratic form in the gradients entering the state space.

To proceed further and determine the conditions satisfying the inequality (\ref{residual-last}), we consider a simple case according to the following ansatz:
\begin{equation}
q^{(A)}_k=0,\qquad A=1,2,\quad k=3,4.
\end{equation}
Under these hypotheses, the residual entropy inequality (\ref{residual-last}) is fulfilled
for all the thermodynamical processes if and only if the following conditions hold true:
\begin{align}
&\varepsilon^{(A)}\equiv\varepsilon^{(A)}(\theta^{(A)}),\qquad A=1,2,\allowdisplaybreaks\\
&\psi^{(A)}_3\frac{\partial \Gamma^{(A)}_0}{\partial \gamma^{(A)}}\geq 0,\qquad A=1,2,\allowdisplaybreaks\label{ineqfinal-2}\\
&{(\rho^{(1)})^2\frac{\partial \Gamma^{(1)}_0}{\partial \rho^{(1)}}-\Gamma^{(1)}_4\frac{\partial \Gamma^{(1)}_0}{\partial \gamma^{(1)}}-\rho^{(1)}\Gamma^{(1)}_0=0,}\allowdisplaybreaks\\
&{\rho^{(2)}\frac{\partial \Gamma^{(2)}_0}{\partial \rho^{(2)}}-\Gamma^{(2)}_0=0,}\allowdisplaybreaks\label{gamma20}\\
&\tau^{(1)}_4\geq 0,\qquad \tau^{(2)}_5 \geq 0,\allowdisplaybreaks\\
&4\tau^{(1)}_4\tau^{(2)}_5\theta^{(1)}\theta^{(2)}-\left(\tau^{(1)}_5 \theta^{(2)}+\tau^{(2)}
_4\theta^{(1)}\right)^2\geq 0,\allowdisplaybreaks\\
&\frac{d\varepsilon^{(A)}}{d\theta^{(A)}}q^{(A)}_A\leq 0,\qquad A=1,2,\allowdisplaybreaks\label{ineqfinal-6}\\
&4q^{(1)}_1q^{(2)}_2(\theta^{(1)})^2(\theta^{(2)})^2\frac{d\varepsilon^{(1)}}{d\theta^{(1)}}\frac{d\varepsilon^{(2)}}{d\theta^{(2)}}-\left(q^{(2)}
_1(\theta^{(1)})^2\frac{d\varepsilon^{(1)}}{d\theta^{(1)}}+q^{(1)}_2(\theta^{(2)})^2\frac{d\varepsilon^{(2)}}{d\theta^{(2)}}\right)^2\ge 0,\allowdisplaybreaks\\
&\psi^{(A)}_3\left(2q^{(A)}_A\frac{d\varepsilon^{(A)}}{d\theta^{(A)}}\frac{\partial \Gamma^{(A)}_0}{\partial\gamma^{(A)}}+(\theta^{(A)})^2\psi^{(A)}_3\left(\frac{\partial \Gamma^{(A)}_0}{\partial\theta^{(A)}}\right)^2\right)\leq 0,\qquad A=1,2,\label{ineqfinal-8}\allowdisplaybreaks\\
&\left(4q^{(1)}_1q^{(2)}_2(\theta^{(1)})^2(\theta^{(2)})^2\frac{d\varepsilon^{(1)}}{d\theta^{(1)}}\frac{d\varepsilon^{(2)}}{d\theta^{(2)}}-\left(q^{(1)}_2(\theta^{(2)})^2\frac{d\varepsilon^{(2)}}{d\theta^{(2)}}+q^{(2)}
_1(\theta^{(1)})^2\frac{d\varepsilon^{(1)}}{d\theta^{(1)}}\right)^2\right)\frac{\partial \Gamma^{(A)}_0}{\partial\gamma^{(A)}}\nonumber\\
&+2q^{(B)}_B\psi^{(A)}_3(\theta^{(A)})^4(\theta^{(B)})^2\frac{d\varepsilon^{(B)}}{d\theta^{(B)}}\left(\frac{\partial \Gamma^{(A)}_0}{\partial\theta^{(A)}}\right)^2\leq 0,\qquad A,B=1,2,\quad A\neq B,\allowdisplaybreaks\\
&\left(\left(4q^{(1)}_1q^{(2)}_2(\theta^{(1)})^2(\theta^{(2)})^2\frac{d\varepsilon^{(1)}}{d\theta^{(1)}}\frac{d\varepsilon^{(2)}}{d\theta^{(2)}}-\left(q^{(1)}_2(\theta^{(2)})^2\frac{d\varepsilon^{(2)}}{d\theta^{(2)}}+q^{(2)}
_1(\theta^{(1)})^2\frac{d\varepsilon^{(1)}}{d\theta^{(1)}}\right)^2\right)\frac{\partial \Gamma^{(1)}_0}{\partial\gamma^{(1)}}\right.\nonumber\\
&+\left.2q^{(2)}_2\psi^{(1)}_3(\theta^{(1)})^4(\theta^{(2)})^2\frac{d\varepsilon^{(2)}}{d\theta^{(2)}}\left(\frac{\partial \Gamma^{(1)}_0}{\partial\theta^{(1)}}\right)^2\right)\frac{\partial \Gamma^{(2)}_0}{\partial \gamma^{(2)}}\nonumber\allowdisplaybreaks\\
&+\psi^{(2)}_3(\theta^{(1)})^2(\theta^{(2)})^4\left(2q^{(1)}_1\frac{d\varepsilon^{(1)}}{d\theta^{(1)}}\frac{\partial \Gamma^{(1)}_0}{\partial\gamma^{(1)}}+\psi^{(1)}_3(\theta^{(1)})^2\left(\frac{\partial \Gamma^{(1)}_0}{\partial\theta^{(1)}}\right)^2\right)\left(\frac{\partial \Gamma^{(2)}_0}{\partial\theta^{(2)}}\right)^2\geq 0.
\end{align}
We note that condition \eqref{gamma20} is equivalent to
\begin{equation}
\Gamma^{(2)}_0=\rho^{(2)}\widetilde{\Gamma}^{(2)}_0(\gamma^{(2)},\theta^{(2)}),
\end{equation}
where $\widetilde{\Gamma}^{(2)}_0$ is a smooth function of the indicated arguments.

The constitutive relations so characterized are worth of some comments. Let us denote with $s_0^{(A)}\equiv s_0^{(A)}(\rho^{(A)},\varepsilon^{(A)})$, with $A=1,2$, the specific entropy at the equilibrium, and consider the relation \eqref{absolute-temp} as an implicit function, 
\[
G(\rho^{(A)},\varepsilon^{(A)}, \theta^{(A)})\equiv\frac{\partial s_0^{(A)}(\rho^{(A)},\varepsilon^{(A)}(\theta^{(A)}))}{\partial\varepsilon^{(A)}}-\frac{1}{\theta^{(A)}}=0; 
\]
then, differentiating the function $G$ with respect to $\rho^{(A)}$, 
we get
\begin{equation}
\label{sign}
\frac{\partial^2 s_0^{(A)}}{\partial\rho^{(A)}\partial\varepsilon^{(A)}}=0,
\end{equation}
\emph{i.e.},
\begin{equation}
s^{(A)}_0(\rho^{(A)},\varepsilon^{(A)})=s^{(A)}_{01}(\varepsilon^{(A)})+s^{(A)}_{02}(\rho^{(A)}),\qquad A=1,2,
\end{equation}
where $s^{(A)}_{01}$ and $s^{(A)}_{02}$ are functions of the indicated arguments.
This relation means that the specific entropy at equilibrium
decomposes in two contributions depending  on the internal energy and the mass density, respectively.
Also, the partial heat fluxes read
\begin{equation}
q^{(A)}=q^{(A)}_1\varepsilon^{(1)}_{,x}+q^{(A)}_2\varepsilon^{(2)}_{,x} = q^{(A)}_1\frac{d\varepsilon^{(1)}}{d\theta^{(1)}}\theta^{(1)}_{,x}+q^{(A)}_2\frac{d\varepsilon^{(2)}}{d\theta^{(2)}}\theta^{(2)}_{,x},
\end{equation}
with $A=1,2$, that describe Fourier-like effects, whereas the entropy flux of the whole blood, taking into account relations \eqref{js} and \eqref{gamma2}, writes
\begin{equation}
\begin{aligned}
J_s=J^{(1)}_s+J^{(2)}_s&=\frac{q^{(1)}}{\theta^{(1)}}+\frac{q^{(2)}}{\theta^{(2)}}-\frac{4}{3}\kappa^{(1)}\frac{(\psi^{(1)}_3)^3}{(\rho^{(1)})^3}\left(\Gamma^{(1)}_4\rho^{(1)}_{,x}+(\rho^{(1)})^2\gamma^{(1)}_{,x}\right)^3\\
&-\frac{4}{3}\kappa^{(2)}(\psi^{(2)}_3)^3(\rho^{(2)})^3(\gamma^{(2)}_{,x})^3.
\end{aligned}
\end{equation}
Finally, since 
\begin{equation}
\frac{d\varepsilon^{(A)}}{d\theta^{(A)}}\geq 0,\qquad A=1,2,
\end{equation}
and taking into account the constraint (\ref{max-entropy}), the inequalities \eqref{ineqfinal-2}, \eqref{ineqfinal-6} and \eqref{ineqfinal-8} imply also
\begin{equation}
\begin{aligned}
&q^{(A)}_A\le 0,\qquad \frac{\partial \Gamma^{(A)}_0}{\partial \gamma^{(A)}}\leq 0,\qquad A=1,2,\\
&2q^{(A)}_A\frac{d\varepsilon^{(A)}}{d\theta^{(A)}}\frac{\partial \Gamma^{(A)}_0}{\partial\gamma^{(A)}}+(\theta^{(A)})^2\psi^{(A)}_3\left(\frac{\partial \Gamma^{(A)}_0}{\partial\theta^{(A)}}\right)^2\geq 0,\qquad A=1,2,\\
\end{aligned}
\end{equation}
that are physically meaningful.
This completes the exploitation of entropy inequality 
for the one-dimensional flow of a blood suspension where the two constituents have 
different temperatures and velocities.

The above procedure allows us to recover partial
Cauchy stress tensors for the RBCs and plasma components which gives some degrees of freedom and so could be specified by using experimental data:
\begin{align}
\label{final-T1}
T^{(1)}&=\tau^{(1)}_0+\tau^{(1)}_1 (\rho^{(1)}_{,x})^2+\tau^{(1)}_2 \rho^{(1)}_{,x}\gamma^{(1)}_{,x}+\tau^{(1)}_3 (\gamma^{(1)}_{,x})^2+\tau^{(1)}_4v^{(1)}_{,x}+\tau^{(1)}_5v^{(2)}_{,x}=\notag\allowdisplaybreaks\\
&=(\rho^{(1)})^2\theta^{(1)} \frac{\partial \psi^{(1)}_0}{\partial \rho^{(1)}}+\theta^{(1)}\frac{(\Gamma^{(1)}_4)^2}{(\rho^{(1)})^3}\left(\rho^{(1)}\frac{\partial \psi^{(1)}_3}{\partial \rho^{(1)}}+2\psi^{(1)}_3\right) (\rho^{(1)}_{,x})^2\notag\allowdisplaybreaks\\
&+2\theta^{(1)}\frac{\Gamma^{(1)}_4}{\rho^{(1)}}\left(\rho^{(1)}\frac{\partial \psi^{(1)}_3}{\partial \rho^{(1)}}+2\psi^{(1)}_3\right) \rho^{(1)}_{,x}\gamma^{(1)}_{,x}\\
&+\theta^{(1)}\rho^{(1)}\left(\rho^{(1)}\frac{\partial \psi^{(1)}_3}{\partial \rho^{(1)}}+2\psi^{(1)}_3\right)(\gamma^{(1)}_{,x})^2+\tau^{(1)}_4v^{(1)}_{,x}+\tau^{(1)}_5v^{(2)}_{,x},\notag\allowdisplaybreaks\\
\label{final-T2}
T^{(2)}&=\tau^{(2)}_0+\tau^{(2)}_3 (\gamma^{(2)}_{,x})^2+\tau^{(2)}_4v^{(1)}_{,x}+\tau^{(2)}_5v^{(2)}_{,x}=\notag\\
&=(\rho^{(2)})^2\theta^{(2)} \frac{\partial \psi^{(2)}_0}{\partial \rho^{(2)}}+\theta^{(2)}\rho^{(2)}\left(\rho^{(2)}\frac{\partial \psi^{(2)}_3}{\partial \rho^{(2)}}+2\psi^{(2)}_3\right)(\gamma^{(2)}_{,x})^2\allowdisplaybreaks\\
&+\tau^{(2)}_4v^{(1)}_{,x}+\tau^{(2)}_5v^{(2)}_{,x}.\notag
\end{align}
We note that the material parameters $\tau^{(A)}_0$ $(A=1,2)$ describe pressure-like effects, whereas the material functions $\tau^{(A)}_4$ and $\tau^{(A)}_5$ assume the role of viscosities, representing the contributions due to the deformations of the single phases of the whole blood. 
 
Both partial Cauchy stress tensors include terms depending on the gradients $\gamma^{(A)}_{,x}$ of the internal variables.

Furthermore, in the expression \eqref{final-T1} of the partial stress tensor $T^{(1)}$, the second term stands for the material parameter linked to the distribution of the red blood cells (recall that it is $\rho^{(1)}=\phi\rho_p$). 

\section{Conclusions}\label{5}
In this paper, a two-component blood model has been formulated. The mathematical model is based on the theory of binary mixtures with two  temperatures and two velocities. The  blood rheology, referring to flow in vessels much larger than the diameter of red blood cells, reflects the behavior of red cells and plasma in bulk of whole blood.  
Our starting point is the framework given in \cite{MasKimAnt}, where general nonlinear constitutive relations  are derived for  partial constitutive quantities of red blood cells and plasma. 
Notice that the governing equations, considering that the two constituents have  different velocities,  involve different material time derivatives; moreover, two independent scalar internal variables, to account for additional dissipative effects, have been introduced. On the contrary, the contribution of external body forces and heat sources have been neglected, as well as momentum and energy exchanges between the components. We plan to investigate some of these aspects in a forthcoming paper.

Since the state space is assumed to include some gradients of the field variables, the constitutive quantities depend also on non-local terms, and the
second law of thermodynamics is exploited by means of the extended Coleman-Noll procedure \cite{CimSelTri}.

All the thermodynamical restrictions determined by the entropy principle on the constitutive equations for a blood suspension with different temperatures, as well as different velocities, are derived and explicitly solved in the one-dimensional case. The solution  provides additional information on the material properties of the model.

It is worth of being stressed that in the blood model here considered,  the expressions of the partial Cauchy stress tensors include viscosity terms; 
differently from the model investigated in \cite{MasKimAnt}, the viscosity coefficients are affected also by the temperature.

Applying the Coleman-Noll extended procedure,  after expanding around the homogeneous equilibrium state the partial Helmholtz free energies, retaining only first order terms in the gradients of some state variables, the partial entropy fluxes have been characterized. They consist of the classical part together to an additional term, the so called extra-flux,  involving cubic terms depending on the gradients of RBCs, plasma mass densities and internal variables.

The theoretical results derived in this paper contain some  phenomenological parameters whose form can be better specialized on the basis of experimental and/or numerical investigations.

Future work will be addressed to the study of the compatibility of non-local constitutive equations with the entropy principle for a blood model in the multi-dimensional case; of course, in this case, general principles of representation theory of vectorial and tensorial
quantities have to be considered.

Furthermore, we plan to  provide  an extension of the blood model  considering nonlinear relations used in non-Newtonian fluid mechanics; for instance, the Reiner-Rivlin or the theory of second-grade fluids.

In this regard,  to get an expression of the Reiner-Rivlin-type stress tensor themodynamically compatible with the second law, it is necessary to extend the state space including  non-localities  of the state variables of higher order.

As a last comment about the constitutive assumption made in \cite{MasKimAnt} for the partial Cauchy stress tensor of RBCs, where also a term involving the squared velocity gradient occurs, we observe that such a term, if included in our model, disappears after the exploitation of entropy inequality, so that in the 
theoretical framework of this paper results incompatible with the second law of thermodynamics.

\section*{Acknowledgements}
Work supported by GNFM-INdAM and by FFABR Unime. 
P.~R. acknowledges the financial support of PRIN 2022 ``Transport phenomena in low dimensional structures: models, simulations and theoretical aspects''.

\end{document}